\documentstyle[12pt]{article}
  \advance\oddsidemargin  by -1in
  \advance\evensidemargin by -1in
  \advance\topmargin      by -0.5in
\makeindex

\def\Pom{{\bf I\!P}}
\def\lsim{\mathrel{\rlap{\lower4pt\hbox{\hskip1pt$\sim$}}
    \raise1pt\hbox{$<$}}}         
\def\gsim{\mathrel{\rlap{\lower4pt\hbox{\hskip1pt$\sim$}}
    \raise1pt\hbox{$>$}}}         
\def\beq{\begin{equation}}
\def\eeq{\end{equation}}
\def\bea{\begin{eqnarray}}
\def\eea{\end{eqnarray}}                  

\begin{document}

\begin{flushright}
{\em ITEP-PH-6/99\\
FZ-IKP(TH)-1999-34}
\end{flushright}

\vspace{1.0cm}
\begin{center}
{\Large \bf How open  charm production and scaling violations probe the 
rightmost hard BFKL pole exchange \\
\vspace{1.0cm}}

{\large \bf N.N. Nikolaev$^{1,2}$ and  V. R. Zoller$^{3}$}\\
\vspace{0.5cm}
$^{1)}${ \em
Institut  f\"ur Kernphysik, Forschungszentrum J\"ulich,
D-52425 J\"ulich, Germany}\\

$^{2)}${\em L.D.Landau Institute for Theoretical Physics, Chernogolovka,
142432 Moscow Region, Russia}\\

$^{3)}${\em Institute for  Theoretical and Experimental Physics,
Moscow 117218, Russia\\
E-mail: zoller@heron.itep.ru} \vspace{0.5cm}\\
{\bf Abstract}
\end{center}
In 1994 Zakharov and the present authors argued that in color dipole (CD) BFKL
approach to DIS excitation of open charm at moderately large $Q^{2}$ is
dominated by hard BFKL exchange. 
In view of the rapid accumulation of the experimental data on small-$x$
charm structure
function of the proton $F_{2}^{c}$ from HERA, we subject the issue of dominance of
the rightmost hard BFKL pole exchange to further scrutiny. Based on CD
BFKL-Regge factorization we report parameter-free predictions for the charm
structure function $F_2^c$ and show that the background to 
the dominant rightmost hard BFKL exchange from
subleading hard BFKL and soft-pomeron exchanges is negligible small from real
photo-production to DIS at $Q^{2}\lsim$50-100 GeV$^{2}$. The agreement with the
experiment is good and lends strong support for the intercept of the rightmost
hard BFKL pole $\Delta_{\Pom}=\alpha_{\Pom}-1=0.4$ as found in 1994 in
the color dipole approach. We comment on the related
 determination of $\Delta_{\Pom}$
from the $x$-dependence of   the longitudinal structure function $F_L(x,Q^2)$ and of
the scaling violation $\partial F_2/\partial \log Q^2$ taken at a suitable value of $Q^2$.
\vspace{0.5cm}\\


\section{Introduction}

In color dipole (CD) approach to small-$x$ DIS excitation of heavy flavor 
is described in terms of interaction of $q\bar{q}$ color
dipoles in the photon with a predominantly small size,
\beq
{ 4 \over Q^{2}+4m_{q}^{2}}\lsim r^{2} \lsim { 1\over m_{q}^{2}}\, ,
\label{eq:1.1}
\eeq
and heavy flavor excitation at large values of the Regge parameter,
\beq
{1\over x}={W^2+Q^2\over4 m^2_c + Q^2}\gg 1\,,
\label{eq:1.2}
\eeq
is an arguably sensitive probe of 
short distance properties of vacuum exchange in QCD. The first analysis of
small-$x$ behavior of
open charm structure function (SF) of the proton $F_2^c$
 in the color
dipole formulation of the Balitsky-Fadin-Kuraev-Lipatov equation \cite{BFKL}
 has been carried out in 1994 
\cite{NZZ94,NZDelta,NZHERA} with an intriguing result that for moderately large 
$Q^{2}$ it is dominated by hard BFKL exchange. As a matter of fact, the
1994 numerical predictions \cite{NZHERA} for $F_2^c$ were 
in the right ball-park and 
agree favorably with the recent experimental data from ZEUS Collaboration 
\cite{ZEUScc}.
Our early observation on hard BFKL dominance in  
\cite{NZZ94,NZDelta} has been based on numerical studies of solutions
of our CD BFKL equation \cite{PISMA1}; more recently this fundamental 
feature of CD
BFKL approach has been related \cite{NZcharm} to nodal properties 
of eigen-functions of
subleading hard BFKL-Regge poles \cite{NZZ97}.
Here we recall that as noticed by Fadin, 
Kuraev and Lipatov in 1975 (\cite{FKL}, see also more detailed discussion by
Lipatov \cite{Lipatov}), incorporation of asymptotic freedom into BFKL
equation  changes the spectrum of the QCD vacuum exchange to series of
isolated BFKL-Regge poles. 

The incorporation of the running coupling and imposition of
the finite range of propagation $R_c$ of perturbative gluons in our
CD BFKL equation provides the interpolation between the BFKL
and DGLAP equations. Although it does not necessarily exhaust all infrared
cutoffs and resummation of higher order corrections, it
emphasizes correctly the principal phenomenon of enhancement of
the infrared region by asymptotic freedom and, confirming the
Lipatov's analysis \cite{Lipatov}, provides the splittings of
the BFKL spectrum into isolated Regge poles and gives the
 subleading BFKL eigenfunctions with expected nodal properties,
 To this end we differ
from recent studies \cite{FL,CC,Brodsky} of NLO corrections
to the original scaling
$\alpha_{S}=const$ and $R_c=\infty$ approximation, in which
the emphasis is still on the scaling approximation, the effects
of finite $R_{c}$ have not been incorporated and the full
resummation to the running coupling has not been yet completed,
for alternative approaches to NLO corrections and infrared effects
see \cite{Thorne,Cudell}.
Our approach is closer to that of Ciafaloni, Catani et al.,
\cite{CCFM} who in their
interpolation between the small-$x$ BFKL dynamics and large-$x$
DGLAP dynamics use running coupling in the manner similar to ours.
True, the incorporation of asymptotic freedom and going beyond the scaling 
approximation makes the intercept 
$\Delta_{0}$ of the leading BFKL pole 
sensitive to the infrared regularization, our $\Delta_{0}=0.4$ 
\cite{NZZ97} must
be regarded as an educated guess; 
the principal emphasis is 
on the nodal properties of subleading
solutions and the dependence of an intercept on the number of nodes.
To this end it is important that the nodes fall into the perturbative region 
of small dipoles and are thus controlled by pQCD better
than the intercept of the leading pole.

Such a discrete spectrum of QCD vacuum exchange has a far-reaching theoretical
and experimental consequences because
the contribution of each isolated hard BFKL pole to scattering amplitudes
and/or SF's would satisfy very powerful 
Regge factorization \cite{Gribov}.
The resulting CD BFKL-Regge factorized expansion allows one to relate in a
parameter-free fashion SF's of different targets, 
$p,\pi,\gamma,\gamma^{*}$ \cite{DER,PION,GamGam} and/or 
contributions of different flavors to 
the proton SF. In this communication we focus on the latter
property of the CD BFKL-Regge factorization and quantify the strength of
the subleading hard BFKL and soft-pomeron background to dominant rightmost
hard BFKL exchange (to be referred to as LHA for the Leading Hard pole
exchange Approximation) improving upon our early somewhat 
simplified application \cite{NZcharm} of the BFKL-Regge factorization to 
$F_2^c$ and extending the analysis to real
photo-production of charm.
We find that this background to LHA is small from real
photo-production to DIS at $Q^{2}\lsim$50-100 GeV$^{2}$. In view of this
fundamental conclusion open charm excitation by real photons and in DIS 
gives a particularly clean access to the intercept of the rightmost hard 
BFKL pole for which our 1994 prediction has been $\Delta_{\Pom}=
\alpha_{\Pom}(0)-1 \approx 0.4$ \cite{NZZ94}.  We show how the soft-pomeron
background dominant at $Q^{2} \lsim $ 5-10 GeV$^{2}$ dies out 
and subleading hard BFKL background builds up for  $Q^{2}\gsim 20$ GeV$^{2}$. 
As one could
have anticipated, because of the small scale (\ref{eq:1.1}) for $c\bar{c}$
color dipoles the soft-pomeron exchange background is negligible small at
all $Q^{2}$ of the practical interest in DIS. Because the CD BFKL-Regge
expansion for color dipole-dipole cross section has already been fixed 
from the related and highly successful phenomenology of light flavor 
contribution to the proton SF the CD BFKL-Regge factorization 
predictions for the charm SF of the proton are parameter 
free. The found nice agreement with the experimental data from ZEUS
Collaboration \cite{ZEUScc} on the charm 
SF of the proton  and open charm
photo-production \cite{DATASIGCC,DATAHERACC} strongly corroborates our 1994 prediction 
$\Delta_{\Pom}=\alpha_{\Pom}(0)-1 \approx 0.4$ for the intercept of the 
rightmost hard BFKL pole. 

Besides charm structure function there are two more observables which are selective
to the dipole size: the longitudinal structure function of the proton $F_L$ 
and  the
scaling violation slope  $\partial F_2/\partial \log Q^2$ \cite{NZDelta}.
 We present the BFKL-Regge factorization results for these observables.
 The recent H1 and ZEUS measurements of scaling violation  do strongly
support $\Delta_{\Pom}= 0.4$ \cite{LOGH1,LOGZ}.


\section{The selectivity of charm structure function to color dipole 
radii}

In color dipole approach DIS at small $x$ is treated in terms of the
interaction of color dipole ${\bf r}$ in the  photon with the color dipole 
${\bf r}_p$ in the target proton which is described by the beam $(b)$, 
target $(t)$ and flavor  independent color dipole-dipole cross section 
$\sigma(x,{\bf r}_b,{\bf r}_t)$. The contribution of excitation of open 
charm to photo-absorption cross section is given by color dipole
factorization formula (we suppress the beam, $\gamma^{*}$, and target, $p$,
subscripts in the cross section)
\beq
\sigma^{c\bar{c}}(x,Q^{2})
=\int dz d^{2}{\bf{r}} dz_{p} d^{2}{\bf{r}_{p}}
|\Psi_{\gamma^*}^{~c\bar{c}}(z,{\bf{r}})|^{2} |\Psi_{p}(z_{p},{\bf{r}_{p}})|^{2}
\sigma(x,{\bf{r},\bf{r^{\prime}}}) =
\int dz d^{2}{\bf{r}} 
|\Psi_{\gamma^*}^{~c\bar{c}}(z,{\bf{r}})|^{2} 
\sigma(x,{\bf{r}})\,.
\label{eq:2.1}
\eeq
where $\sigma(x,{\bf{r}})$ stands for interaction of the beam dipole with
the target nucleon. Here $|\Psi_{\gamma^*}^{~c\bar{c}}(z,{\bf{r}})|^{2}$ is
a probability to find in the photon the $c\bar{c}$ color dipole with the
charmed quark carrying fraction $z$ of the photon's light-cone momentum.
The well known result of \cite{NZ91} for the transverse (T) and longitudinal 
(L)  photons is 
\beq
|\Psi_T^{c\bar c}(z,r)|^2={2\alpha_{em}\over 3\pi^2}
\left\{\left[z^2+(1-z)^2\right]\varepsilon^2K_1(\varepsilon r)^2
+m_c^2K_0(\varepsilon r)^2\right\},
\label{eq:2.2}    
\eeq
\beq
|\Psi_L^{c\bar c}(z,r)|^2={8\alpha_{em}\over 3\pi^2}
Q^2z^2(1-z)^2K_0(\varepsilon r)^2\,,
\label{eq:2.3}    
\eeq
where $K_{0,1}(y)$ are the modified Bessel functions,  $\varepsilon^2=z(1-z)Q^2+m_c^2$
and $m_c=1.5\,{\rm GeV}$ is the $c$-quark mass. Hereafter we focus on the charm structure
function  
\bea
F_2^{c}(x_{Bj},Q^2)=
{Q^2\over {4\pi^2\alpha_{em}}}\int_0^1 dz\int
d^2\vec r\left[|\Psi_L^{c\bar c}(z,r)|^2+|\Psi_T^{c\bar c}(z,r)|^2\right]\sigma(x,r)
\nonumber \\
=\int{ d r^{2}\over r^{2}}{\sigma(x,r) \over r^{2}} W_{2}(Q^{2},m_{c}^{2},r^2)
\,.
\label{eq:2.4}
\eea
We present the results for $F_2^{c}$ as a function of the conventional 
Bjorken variable $x_{Bj}$, for the relationship between the Regge parameter
$x$ and the Bjorken variable see eq.~(\ref{eq:3.4}) below.

The Bessel function $K_{1}(y)$ has the $\sim {1\over y}$ singularity at $y\to 0$
and decreases exponentially at $y\gsim 1$, i.e., for color dipole 
\beq
r \gsim {1\over \varepsilon}\,,
\label{eq:2.5}
\eeq
cf. eq.~(\ref{eq:1.1}). However, because for small dipoles $\sigma(x,r)\sim r^{2}$, 
the dipole size integration in (\ref{eq:2.4}) is well convergent at small $r$. A
detailed  analysis of  the weight function $W_{2}(Q^{2},m_{c}^{2},r^2)$ found upon 
the $z$ integration has been carried out in \cite{NZDelta,NZHERA}, we only cite
the principal results: (i) at moderate $Q^{2} \lsim 4m_{c}^{2}$ the weight
function has a peak at $r \sim {1\over m_{c}}$, (ii) at very high $Q^{2}$ the peak 
develops a plateau for dipole sizes in the interval (\ref{eq:1.1}).
One can say that for moderately large $Q^{2}$ excitation of open charm 
probes (scans) the dipole cross section at a special dipole size $r_{S}$
(the scanning radius) 
\beq
r_{S} \sim {1\over m_{c}}\,.
\label{eq:2.6}
\eeq
The difference from light flavors is that in contrast to the peak for heavy
charm the $W_{2}$ for light flavors always has a broad plateau which extends 
up to large dipoles $r \sim {1\over m_{q}}$.


\section{Scanning radius and nodes of subleading CD BFKL eigen-cross sections
and eigen-structure functions} 

In the Regge region of ${1\over x} \gg 1$  CD cross section $\sigma(x,r)$
satisfies the CD BFKL equation
\beq
{\partial \sigma(x,r) \over \partial \log{1\over x}}={\cal K}\otimes \sigma(x,r)\,,
\label{eq:3.1}
\eeq
for the kernel ${\cal K}$ of CD approach see \cite{PISMA1}.
The solutions with Regge behavior
\beq
\sigma_{m}(x,r)=
\sigma_{m}(r)\left({1\over x}\right)^{\Delta_{m}}
\label{eq:3.0} 
\eeq
satisfy the eigen-value 
problem
\beq
{\cal K}\otimes \sigma_{m}=\Delta_{m}\sigma_{m}(r)\,
\label{eq:3.2}
\eeq
and the CD BFKL-Regge expansion for beam-target symmetric 
color dipole-dipole cross section
reads \cite{NZZ94,PION}
\beq
\sigma(x, r,r_p)=\sum_{m=1} C_m 
\sigma_m(r)\sigma_m(r_p)\left({x_0\over x}\right)^{\Delta_m}.
\label{eq:3.3}
\eeq
 The practical calculation of $\sigma(x,r,r_p)$
    requires the boundary condition $\sigma(x_0,r,r_p)$ 
at certain $x_0\ll 1$.
We take for boundary condition at $x=x_0$ the Born approximation,
$$\sigma(x_0,r,r_p)=\sigma_{Born}(r,r_p)\,,$$ 
 i.e. evaluate dipole-dipole scattering via the two-gluon exchange.
This leaves the starting point $x_0$ the sole parameter.
We follow the choice $x_0=0.03$ which  met  with  remarkable
 phenomenological success 
\cite{NZZ97,DER,PION}.

Here one should not confuse $x$ in the definition of the Regge parameter
(\ref{eq:1.2}) with the Bjorken variable
\beq
x_{Bj}={Q^{2}\over W^{2}+Q^{2}}= x\cdot {Q^{2} \over Q^{2}+4m_{c}^{2}}\, .
\label{eq:3.4}
\eeq
Our choice of normalization of eigen-functions $\sigma_m(r)$ is such that
upon calculation of the expectation value over the target proton dipole
distribution in (\ref{eq:2.1}) 
\beq
\sigma(x, r)=\sum_m \sigma_m(r)\left({x_0\over x}\right)^{\Delta_m}.
\label{eq:3.5}
\eeq

The properties of our CD BFKL equation and the choice of
physics motivated boundary condition were discussed in detail elsewhere 
\cite{NZDelta,NZHERA,NZcharm,NZZ97,DER}, 
here we only recapitulate features 
relevant to the considered problem. Incorporation of asymptotic freedom 
exacerbates well known infrared sensitivity of
the BFKL equation and infrared regularization by infrared freezing of the 
running coupling $\alpha_S(r)$ and modeling of confinement of gluons 
by the finite propagation radius of perturbative gluons $R_c$ need to be
 invoked.

The leading eigen-function $\sigma_0(r)$
for ground state. i.e., for the rightmost hard BFKL pole is node free.  
The  subleading  eigen-function for excited state $\sigma_m(r)$ has $m$ nodes. 
We find $\sigma_m(r)$ numerically \cite{NZZ97,DER}, for the semi-classical 
analysis see Lipatov \cite{Lipatov}. The intercepts (binding energies) follow
to a good approximation the law $\Delta_{m}= \Delta_{0}/(m+1)$. For the 
preferred 
$R_c=0.27\, {\rm fm}$ as chosen in 1994 in \cite{NZHERA,NZDelta} and supported by
recent analysis \cite{MEGGI} of lattice QCD data 
we find $\Delta_{0}=\Delta_{\Pom}=0.4$,  
the  node of $\sigma_{1}(r)$ is located at $r=r_1\simeq 
0.056\,{\rm fm}$, for larger $m$ the rightmost node moves to a somewhat  
larger $r=r_1\sim 0.1\, {\rm fm}$. The second node of eigen-functions with
$m= 2,3$ is located at  $r_{2}\sim 3\cdot 10^{-3}~ {\rm fm}$ which corresponds
to the momentum transfer scale 
$Q^{2} ={1\over r_{2}^{2}}=5\cdot 10^{3}$ GeV$^{2}$.
The third node of $\sigma_{3}(r)$ is located at $r$ beyond the reach of any
feasible DIS experiments. It has been found \cite{NZZ97} that the BFKL-Regge
 expansion (\ref{eq:3.6}) truncated at $m=2$ appears to be very successful
 in describing of the proton SF's at 
  $Q^2\lsim 200$ GeV$^2$. However, at higher $Q^2$ and
moderately
small $x\sim x_0=0.03$ the background of the CD BFKL solutions with
 smaller intercepts ($\Delta_m < 0.1$)
should be taken into account (see below).

The exchange by perturbative gluons is a dominant mechanism for small dipoles
$r\lsim R_c$. In Ref.\cite{NZHERA} interaction of large dipoles
has been modeled by the non-perturbative, soft mechanism with
intercept $\alpha_{\rm soft}(0)-1=\Delta_{\rm soft}=0$ i.e. flat vs. 
$x$ at small $x$.  The
exchange by two non-perturbative gluons has been behind the
specific parameterization of $\sigma_{\rm soft}(r)$ suggested in \cite{JETPVM}
and used later on in \cite{NZcharm,NZZ97,DER,PION,GamGam} and here, 
see also Appendix.

Via equation (\ref{eq:2.4}) each hard CD BFKL eigen-cross section plus soft-pomeron 
CD cross section defines the corresponding
eigen-SF  $f_m^{c}(Q^2)$ and we arrive at the CD BFKL-Regge
expansion for the charm SF of the proton 
$(m={\rm soft},0,1,..)$
\beq
F_2^{c}(x_{Bj},Q^2)=\sum_{m} f_m^{c}(Q^2)\left(x_0\over x\right)^{\Delta_m}\,,
\label{eq:3.6}
\eeq
Now comes the crucial observation that numerically 
$r_{1} \sim {1\over 2} r_{S}$ 
and the node of hard CD BFKL eigen-cross sections is located within the peak 
of the weight function $W_{2}$. Consequently, in the calculation of open 
charm eigen-SFs $f_m^{c}(Q^2)$ one scans the eigen-cross 
section in the vicinity of the node, which leads to a strong suppression of 
subleading $f_m^{c}(Q^2)$. This point is illustrated in fig.~1 in which the 
subleading BFKL-to-rightmost BFKL and soft-pomeron-to-rightmost BFKL ratio 
of eigen-SFs is shown. For the charm quark mass which is the
sole new parameter we take 
$m_{c}=1.5$ GeV. Because for charm the weight function
is peaked at $r\approx r_{s}$ and, in contrast to that for light flavors,
does not extend to larger $r$, the hierarchy and nodal structure of charm 
eigen-SFs $f_m^{c}(Q^2)$ differs substantially from that
for light flavors discussed in \cite{NZZ97}: (i) the node of $f_1^{c}(Q^2)$ 
shifts from $Q_{1}^{2} \approx 60 $GeV$^2$ down to
 $Q_{1}^{2} \approx 20 $GeV$^2$, 
(ii) the first
node of $f_2^{c}(Q^2)$ shifts from $Q_{1}^{2} \approx 30 $GeV$^2$ down 
to $Q_{1}^{2} \approx 1 $GeV$^2$ and $f_2^{c}(Q^2)\sim 0$ up to $Q \lsim 
20 $GeV$^2$, (iii)the background SF 
 $f_3^{c}(Q^2)$ is free of the first node at $Q_{1}^{2}\sim 
20$GeV$^2$ which is present in eigen-SF for light flavors.


\section{Predictions from CD BFKL-Regge factorization for open charm 
structure function and photoproduction}

The results shown in fig.~1 form the basis of the CD BFKL-Regge phenomenology
of open charm production. 
Because a probability to find large color dipoles in the photon decreases
rapidly with the quark mass, the contribution from soft-pomeron exchange
to open charm excitation is very small down to $Q^{2}=0$. In contrast to
that for light flavors soft-pomeron exchange was the dominant mechanism 
at small $Q^{2}$, see \cite{PION}. Large color dipoles are present in
the photon and keep contributing to $F_2^c$ even for very 
large $Q^{2}$ but relevance of soft-pomeron exchange diminishes 
gradually with $Q^{2}$. 
As we discussed elsewhere \cite{PION}, for still higher solutions, $m\geq 3$, 
all intercepts are very small anyway, $\Delta_m\ll \Delta_{0}$,
 For
this reason, for the purposes of practical phenomenology we can truncate 
expansion (\ref{eq:3.6}) at $m=3$ lumping in the term $m=3$ contributions 
of still higher singularities with $m\geq 3$. The term $m=3$
which is a combination
of higher CD BFKL solutions, 
\beq
\sigma_3(r)=\sigma_{Born}(r)-\sum_{m=0}^2\sigma_m(r)\,,
\label{eq:SIGMA3}
\eeq
 is  endowed
 with the effective intercept $\Delta_3=0.06$
 and is presented in Appendix
 in its analytical form. Introducing such a term extends the applicability
 region of the truncated CD BFKL-Regge expansion up to
 $Q^2\sim 10^4-10^5$ GeV$^2$.
  Notice that in \cite{PION} we accounted for 
the $m\geq 3$-background in a somewhat different way 
 than that accepted here.
 However,
 the difference between two approaches becomes substantial only at 
$Q^2\gsim 300\, GeV^2$,
  and do not
affect the numerical results at smaller $Q^2$
 (mind the Regge suppression factor $(x/x_0)^{\Delta_0-\Delta_m}$).    

 As fig.~1 shows, the hierarchy
of $f_{m}^{c}(Q^{2})$ is exceptional in that in the very broad of
$Q^{2}$ of the practical interest the contribution from $m=2$ is negligible
small compared to the contribution from $m=3$. For this reason the term
$m=3$ is numerically  important for description of charm structure 
function at $Q^{2} \gsim 50$ GeV$^{2}$. This hierarchy of 
$f_{m}^{c}(Q^{2})$ has been overlooked in our early analysis 
\cite{NZcharm} where the truncation of the BFKL-Regge expansion
at $m\leq 2$ has been made.

Color dipole cross section is flavor independent and the charm quark mass 
$m_{c}=1.5$ GeV is the sole new parameter in our predictions 
from the CD BFKL-Regge factorization for 
open charm SF of the proton presented in fig.~2 
as a function of the Bjorken variable $x_{Bj}$ and the results for
open charm photoproduction shown in fig.~3. As eq.~(\ref{eq:3.4}) shows, for
small $Q^{2}$ the starting point $x_{0}=3\cdot 10^{-2}$ of the BFKL
evolution corresponds to progressively smaller $x_{Bj}$ and the CD
BFKL-Regge expansion is applicable at 
\beq
x_{Bj} \leq  x_{0}\cdot {Q^{2} \over Q^{2}+4m_{c}^{2}}\, .
\label{eq:4.4}
\eeq
and in real photoproduction at 
\beq
\nu \geq \nu_{0} ={2m_{c}^{2} \over m_{p}x_{0}} \sim 150\, {\rm GeV}\, .
\label{eq:4.5}
\eeq
In order to give a crude idea on finite-energy effects at large $x_{Bj}$
and not so large values of the Regge parameter we stretch the theoretical
curves a bit to $x\gsim x_{0}$ and/or lower energies $\nu\lsim \nu_{0}$ 
multiplying the BFKL-Regge expansion result (\ref{eq:3.6}) by the 
purely phenomenological factor $(1-x)^{5}$ motivated by the familiar 
behavior of the gluon SF of the proton $G(x_{Bj}) 
\sim (1-x_{Bj})^{n}$ with the exponent $n\sim 5$. 

We comment first on the results on $F_2^c$. The solid curve 
is a result of the complete CD BFKL-Regge expansion. The long-dashed 
curve is the pure rightmost hard BFKL pomeron contribution (LHA). 
The soft (S) pomeron exchange 
contribution is numerically too small to be shown separately. The
sum of the rightmost hard BFKL (LH for the Leading Hard) and soft pomeron
exchanges (LHSA) is shown by the dotted curve in the box for $Q^{2}=4$ GeV$^2$
and practically merges with the curve for complete CD BFKL-Regge expansion.
This is not unexpected from fig.~1 which shows that for $Q^{2} \lsim 10$ 
GeV$^{2}$ there is a strong cancellation between soft and  subleading 
contributions with $m=1$ and $m=3$. Consequently, for this dynamical reason 
in this region of $Q^{2} \lsim10$ GeV$^{2}$ we have an effective one-pole 
picture and LHA gives reasonable description of  $F_2^c$.
  In agreement with the nodal structure of subleading
eigen-SFs LHA over-predicts slightly 
 $F_2^c$ at $Q^{2} \gsim 
30$ GeV$^{2}$, where the negative valued subleading hard BFKL exchanges
overtake the soft-pomeron exchange, see fig.~1, and the background from 
subleading hard BFKL exchanges becomes substantial at $Q^{2} \gsim 30$ GeV$^2$
and even the dominant component of $F_2^c$ at $Q^{2}\gsim 200$
GeV$^{2}$ and $x\gsim 10^{-2}$. In this region of $Q^{2}$ the soft-pomeron
exchange is numerically so small the curves for LHSA and LHA merge with
each other within the thickness of curves and the LHSA curves are omitted.
We predict that open charm SF
is dominated entirely by the contribution from the rightmost hard BFKL pole
at $Q^{2} \lsim 20$ GeV$^{2}$, which is due to strong cancellations 
between the soft-pomeron and subleading hard BFKL exchanges, see fig.~1.
The soft-subleading cancellations become less accurate at smaller $x$, but
at smaller $x$ the both soft and subleading hard BFKL exchange become 
rapidly Regge suppressed $\propto x^{\Delta_{\Pom}}, 
~x^{{1\over 2}\Delta_{\Pom}}$, respectively. 

In fig.~2 we compare our CD BFKL-Regge predictions for  small-$x$ 
charm SF of the proton shown by the solid curve to the 
recent experimental data from the ZEUS Collaboration \cite{ZEUScc} and 
find very good agreement between theory and experiment which
lends support to our 1994 evaluation $\Delta_{\Pom}=0.4$ of the 
intercept of the rightmost hard BFKL pole in the color dipole approach 
with running strong coupling. The negative 
valued contribution from subleading hard BFKL exchange is important for
bringing the theory to agreement with the experiment at large $Q^{2}$.
Very recently Donnachie \& Landshoff have parameterized the same ZEUS data 
in terms of the two-pole (soft+hard) Regge model \cite{DLcharm} and 
concluded that they are consistent with dominance of the pure
hard pole exchange with $\Delta \approx 0.44$. Our dynamical model has more 
predictive power because it quantifies corrections to single-pole
dominance. For instance, it predicts unequivocally that single-pole 
approximation would break at $Q^{2}\gsim 50$ GeV$^2$. It also predicts
that the background to the rightmost hard BFKL pole with $\Delta_{\Pom}=0.4$
changes from the negligible small value at small $Q^{2}$ to
negative-valued subleading, $m=3$, hard background BFKL exchange with the 
 intercept
 $\Delta_{3}=0.06 
\approx {1\over 7}\Delta_{\Pom}$ with a weak for $Q^{2}$ of the practical 
interest admixture from subleading, $m=1$ \& $m=2$, 
exchanges with larger intercepts  $\Delta_{1}=0.22 $ and $\Delta_{2}=0.15$.
If one would make the effective single-pole fits of the form 
$F_2^c \propto \left({1\over x}\right)^{\Delta_{eff}}$, then according
to our approach $\Delta_{eff} \sim \Delta_{\Pom}=0.4$ for
$Q^{2}\lsim 20 $ GeV$^{2}$ and $\Delta_{eff} \gsim \Delta_{\Pom}$ and
would rise gradually for $Q^{2} \gsim 20$ GeV$^{2}$.

In fig.~3 we compare our predictions from CD BFKL-Regge factorization
for real photoproduction of open charm with the experimental data
from fixed target \cite{DATASIGCC} and HERA collider H1 and ZEUS 
\cite{DATAHERACC} experiments. The legend of theoretical curves is the
same as in fig.~2: the solid curve is a result of the complete 
BFKL-Regge expansion, the dotted curve is for the Leading Hard $+$ Soft 
exchange Approximation (LHSA), the long-dashed curve is the pure
rightmost hard BFKL pomeron contribution (LHA). The fixed target data
are in the region of moderately large Regge parameter when finite-$x$
corrections modeled by the factor $(1-x)^5$ show up. The agreement 
between theory and experiment is good and must be regarded as an important 
confirmation of $\Delta_{\Pom}=0.4$ for the rightmost hard BFKL exchange.
 For an alternative interpretation of charm photoproduction
see \cite{TOLYA}).

\section{Determination of the pomeron intercept $\Delta_{\Pom}$ from
measurements of $F_L(x,Q^2)$ and $\partial F_2/\partial \log Q^2$}

It has been demonstrated in \cite{NZDelta} that the longitudinal structure
function $F_L(x,Q^2)$ and the slope of the structure function
$\partial F_T/\partial \log Q^2$ emerge as local probes of the dipole
cross section at $r^2\simeq 11./Q^2$ and $r^2\simeq 2.3/Q^2$, respectively.
The subleading CD BFKL cross sections have their rightmost node
at $r_1\sim 0.05-0.1$ fm. Therefore, one can zoom at the leading CD BFKL  pole
contribution and measure the pomeron intercept
$\Delta_{\Pom}$ from the $x$-dependence of $F_L(x,Q^2)$
at $Q^2\sim 10-30$ GeV$^2$ and of  $\partial F_2/\partial \log Q^2$ at
$Q^2\sim 2-10$ GeV$^2$.

In Fig.4 we show the ratio $f_{\rm Lm}/f_{\rm L0}$ of subleading to leading
longitudinal eigen-SF and soft to leading eigen-SF.
From Fig.4   it follows that in the CD BFKL-Regge expansion
for $F_L$ (see Appendix)
 the discussed above cancellation of the  soft-subleading contributions
 is nearly exact at $Q^2\sim 10-30$ GeV$^2$. This results in the leading hard pole
 dominance in this region (see the box  $Q^2=20$ GeV$^2$ in Fig.5).

In Fig.6 we presented the ratio $d_m(Q^2)/d_0(Q^2)$
  of logarithmic derivatives
$d_m(Q^2)=\partial f_m(Q^2)/\partial \log Q^2$ of the all flavor eigen-SF
 for $m={\rm soft},0,1,2,3$ (see \cite{DER} for more details).
The pattern of cancellations of the soft-subleading contributions
 is somewhat different in this case and we predict that
 the leading hard pole dominates the region of several GeV$^2$.
 The $x_{Bj}$-dependence of the log-derivative 
 $\partial F_2/\partial \log Q^2$  is shown in Fig.7
for $Q^2=0.75,5$ and $40$ GeV$^2$.  
The cancellation is  exact in the case of $Q^2\simeq 4$ GeV$^2$.  
Comparison with preliminary HERA data \cite{LOGH1,LOGZ} exhibits good agreement 
of our calculations with experiment.


\section{Conclusions}

Color dipole approach to the BFKL dynamics predicts uniquely decoupling 
of subleading hard BFKL exchanges from open charm SF of
the proton at $Q^2\lsim 20\,{\rm GeV^2}$, from $F_L$ at $Q^2\simeq 20\,{\rm GeV^2}$ 
and from $\partial F_2/\partial \log Q^2$ at 
 $Q^2\simeq 4\,{\rm GeV^2}$. This decoupling is due to
dynamical cancellations between contributions of different subleading
hard BFKL poles and leaves us with an effective soft+rightmost hard BFKL 
two-pole approximation with intercept of the 
soft pomeron $\Delta_{{\rm soft}}=0$. 
We predict strong cancellation between the soft-pomeron 
and subleading hard BFKL contribution to $F_2^c$ in the experimentally
 interesting region
of $ Q^{2} \lsim 20$ GeV$^{2}$, in which $F_2^c$
is dominated entirely by the contribution from the rightmost hard BFKL pole.
This makes open charm in DIS at $Q^2\lsim 20$ GeV$^{2}$ a unique 
handle on the intercept of the rightmost hard BFKL exchange.
Similar hard BFKL pole dominance holds for $F_L(x,Q^2)$ 
and  $\partial F_2/\partial \log Q^2$.  
At still higher values of $Q^{2}$ the soft-pomeron exchange is predicted to
die out and negative valued background contribution from subleading hard BFKL
exchange with effective  intercept $\Delta_{3}\approx 0.06$ 
becomes substantial at not too small $x\sim x_0$. The agreement with the presently available
experimental data on open charm in DIS and real photoproduction and the recent data
on scaling violation $\partial F_2/\partial \log Q^2$ is good
and confirms the CD BFKL prediction of the intercept $\Delta_{\Pom}=0.4$ 
for the rightmost hard BFKL-Regge pole. The experimental 
confirmation of our predictions for hierarchy 
of soft-hard exchanges as function of $Q^{2}$ 
would be a strong argument in favor of the CD BFKL
approach. \\

{\bf Acknowledgments: } This work was partly supported by the grants
INTAS-96-597, INTAS-97-30494 and DFG 436RUS17/11/99.

\newpage

\section{ Appendix}
\subsection{CD BFKL charm eigen-SF}

The shape and nodal properties of eigen-functions $\sigma_m(r)$ as 
a function of $r$ and/or eigen-SFs $f^{c}_m(Q^2)$ 
as a function of $Q^2$ is well understood \cite{NZZ97,DER,NZcharm}.
However, the  eigen-cross sections $\sigma_m(r)$ are only  available
as a numerical solution to the running color dipole BFKL equation.
On the other hand, for the practical  applications it is convenient to 
have analytical parameterization for eigen-SFs
$f^{c}_m(Q^2)$, which for the rightmost hard BFKL pole is of the form
\beq
f^c_0(Q^2)=
a_0{R_0^2Q^2\over{1+ R_0^2Q^2 }}
\left[1+c_0\log(1+r_0^2Q^2)\right]^{\gamma_0}\,,
\label{eq:F20C}
\eeq
where $\gamma_0=4/(3\Delta_0)$, while for the subleading hard BFKL poles
\beq
f^c_m(Q^2)=a_m f_0(Q^2){1+ K_m^2Q^2\over{1+ R_m^2Q^2 }}
\prod ^{m_{max}}_{i=1}\left(1-{z\over z^{(i)}_m}\right)\,,\,\, m\geq 1\,,
\label{eq:FNC}
\eeq
where $m_{max}=$min$\{m,2\}$
and
\beq
z=\left[1+c_m\log(1+r_m^2Q^2)\right]^{\gamma_m}-1 ,\,\,\,
\gamma_m=\gamma_0 \delta_m.
\label{eq:ZFNC}
\eeq
The parameters  tuned to reproduce  the numerical results for 
$f^{c}_m(Q^2)$ at $Q^2\lsim 10^4\, GeV^2$ are listed in the Table 1.

The soft component of the charm SF as derived from
$\sigma_{\rm soft}(r)$ taken from \cite{JETPVM} is parameterized as 
\beq
f^{c}_ {\rm soft}(Q^2)= 
{a_{\rm soft}R^2_{\rm soft}Q^2\over{1+R^2_{\rm soft} Q^2 }}
\left[1+c_{\rm soft}\log(1+r^2_{\rm soft}Q^2)\right]\,,
\label{eq:FCSOFT}
\eeq
with parameters cited in the Table 1.

\vspace{5mm}

\begin{center}
Table 1. CD BFKL-Regge charm structure functions parameters.
\begin{tabular}{|c|c|c|c|c|c|c|c|c|} \hline
$m$& $a_m$ &  $c_m$ & $r_m^2,$         & $ R_m^2,$       &$ K^2_m,$  &$z^{(1)}_m$ & $z^{(2)}_m$ & $\delta_m$ \\ 
   &       &        & ${\rm GeV^{-2}}$ & ${\rm GeV^{-2}}$&${\rm GeV^{-2}}$ &            &             &     \\ \cline{1-9}
0  &  0.02140        & 0.2619  & 0.3239 & 0.2846&          &        &  & 1. \\ \cline{1-9}
1  & 0.0782         & 0.03517  &0.0793  &0.2958 & 0.2846  &  0.2499&  & 1.9249\\ \cline{1-9}
2  & 0.00438        &0.03625   &0.0884  &0.2896 & 0.2846  &  0.0175 &3.447&1.7985\\ \cline{1-9}
3  &$-0.26313$      &2.1431    &$3.7424\cdot 10^{-2}$ &$8.1639\cdot 10^{-2}$ & 0.13087   &158.52   & 559.50 &0.62563  \\ \cline{1-9}
{\rm soft}& 0.01105  &0.3044   &0.09145 &0.1303 &         &         &     &        \\ \cline{1-9}
\end{tabular} 
\end{center}
\vspace{5mm}

\subsection{CD BFKL longitudinal eigen-SF}

The CD BFKL expansion for the vacuum component of the  all flavor longitudinal
 SF
reads
\beq
F_L(x_{Bj},Q^2)=\sum_{m}  f_{Lm}^{uds}(Q^2)
\left(x_0\over x^{\prime}\right)^{\Delta_m}+\sum_{m} f_{Lm}^{c}(Q^2)
\left(x_0\over x\right)^{\Delta_m}\,,
\label{eq:FLRegge}
\eeq
where  $1/x$ for the charm SF is specified by  eq.(\ref{eq:1.2})
 and  the light flavor Regge
 parameter is
\beq
{x_0\over x^{\prime}}= x_0{{Q^2+W^2}\over {Q^2+m_{\rho}^2}}\,,
\label{eq:XREGGE}
\eeq
 where
$m_{\rho}$ is the $\rho$-meson mass \cite{DER}.
The parameterizations for the all flavor longitudinal
eigen-SFs $f_{Lm}(Q^2)$ and the longitudinal charm
eigen-SFs $f^c_{Lm}(Q^2)$ related to the light flavor
 eigen-SFs as  $f^{uds}_{Lm}=f_{Lm}-f^c_{Lm}$ are presented here.
For the all flavor longitudinal eigen-structure functions we have
\beq
f_{\rm L0}(Q^2)=
a_{\rm 0}{R_{\rm 0}^2Q^2\over{1+ R_{\rm 0}^2Q^2 }}{K_{\rm 0}^2Q^2\over{1+ K_{\rm 0}^2Q^2 }}
\left[1+c_{\rm 0}\log(1+r_{\rm 0}^2Q^2)\right]^{\gamma_0}\,,
\label{eq:FL0}
\eeq
where  $\gamma_0={4\over {3\Delta_0}}$, $R_{\rm 0}^2=13.742\, {\rm GeV^{-2}}$,  $K_{\rm 0}^2=0.72578\, {\rm GeV^{-2}}$
 and 
\beq
f_{\rm Lm}(Q^2)=a_{\rm m} f_{\rm L0}(Q^2)
\prod ^{m}_{i=1}\left(1-{z\over z^{(i)}_{\rm m}}\right)\,,\,\, m\geq 1\,,
\label{eq:FLN}
\eeq
where
\beq
z=\left[1+c_{\rm m}\log(1+r_{\rm m}^2Q^2)\right]^{\gamma_{\rm m}}-1 ,\,\,\,
\gamma_{\rm m}=\gamma_0 \delta_{\rm m}\,.
\label{eq:ZLFN}
\eeq
The parameters  adjusted to reproduce the numerical results for 
$f_{\rm L\,m}(Q^2)$ at $Q^2\lsim 10^4\, GeV^2$ are listed in the Table 2.

The soft component of the longitudinal structure function  is parameterized as 
\beq
f_ {\rm L\,soft}(Q^2)= 
a_{\rm soft}\left({r^2_{\rm soft}Q^2\over{1+r^2_{\rm soft} Q^2 }}\right)^2
{1+R^2_{\rm soft}Q^2\over 1+K^2_{\rm soft}Q^2}\,,
\label{eq:FLSOFT}
\eeq
with  $R^2_{\rm soft}=0.17374\,GeV^{-2}$, 
$ K^2_{\rm soft}=0.61476\,GeV^{-2}$ and $a_{\rm soft}$,
$r^2_{\rm soft}$
 cited in the Table 2.

\vspace{0.5cm}

\begin{center}
Table 2. CD BFKL-Regge longitudinal structure functions parameters.
\begin{tabular}{|l|l|l|l|l|l|l|l|l|} \hline
${\rm m}$ & $a_{\rm m}$  & $c_{\rm m}$ & $r_{\rm m}^2$ & 
 $z^{(1)}_{\rm m}$ & $z^{(2)}_{\rm m}$ & $z^{(3)}_{\rm m}$ & $\delta_{\rm m}$ & $\Delta_{\rm m}$ \\ 
  &                     &                      & ${\rm GeV^{-2}}$  &              &         &        &        &       \\ \cline{1-9}
0 & 9.756$\cdot 10^{-3}$& 0.24835              &0.5193   &                        &         &        & 1.     & 0.402 \\ \cline{1-9}
1 & 0.34897            &3.5370$\cdot 10^{-2}$ &9.6065   &  4.5613                &         &        & 2.5472 &  0.220 \\ \cline{1-9}
2 & 0.27132             &1.8934$\cdot 10^{-2}$ &5.8656   &   1.9627               &12.172   &        & 3.7111 &  0.148  \\ \cline{1-9}
3 & 2.38323             &2.3467$\cdot 10^{-3}$ &5.1690   &  $ 7.2783\cdot 10^{-2}$&0.20309  &0.33768 & 2.6115 &  0.06 \\ \cline{1-9}
{\rm soft}&0.03181      &                      &7.8172   &                        &         &        &        &  0.   \\  \cline{1-9}
\end{tabular}
\end{center}
\vspace{0.5cm}

\subsection{CD BFKL longitudinal charm eigen-SF}

For the longitudinal charm eigen-structure functions 
the parameterization reads
\beq
f^{c}_{\rm L0}(Q^2)=
a_{\rm 0}{R_{\rm 0}^2Q^2\over{1+ R_{\rm 0}^2Q^2 }}{K_{\rm 0}^2Q^2\over{1+ K_{\rm 0}^2Q^2 }}
\left[1+c_{\rm 0}\log(1+r_{\rm 0}^2Q^2)\right]^{\gamma_0}\,,
\label{eq:FL0C}
\eeq
where  $\gamma_0={4\over {3\Delta_0}}$    and 
\beq
f^{c}_{\rm Lm}(Q^2)=a_{\rm m} f^{c}_{\rm L0}(Q^2){{1+R^2_m Q^2}\over {1+K^2_m Q^2}}
\prod ^{m_{max}}_{i=1}\left(1-{z\over z^{(i)}_{\rm m}}\right)\,,\,\, m\geq 1\,,
\label{eq:FLNC}
\eeq
where $m_{max}={\rm min}\{2,m\}$,
\beq
z=\left[1+c_{\rm m}\log(1+r_{\rm m}^2Q^2)\right]^{\gamma_{\rm m}}-1 ,\,\,\, 
\gamma_{\rm m}=\gamma_0 \delta_{\rm m}\,. 
\label{eq:ZLFNC}
\eeq
The parameters  adjusted to reproduce the numerical results for 
$f^c_{\rm L\,m}(Q^2)$ at $Q^2\lsim 10^4$ GeV$^2$ are listed in the Table 3.

The soft component of the longitudinal  charm structure function  is parameterized as 
\beq
f^{c}_ {\rm L\, soft}(Q^2)= 
a_{\rm soft}\left({r^2_{\rm soft}Q^2\over{1+r^2_{\rm soft} Q^2}}\right)^2 \,,
\label{eq:FLCSOFT}
\eeq
with parameters cited in the Table 3.

\vspace{0.5cm}

\begin{center}
Table 3. CD BFKL-Regge longitudinal charm structure functions parameters.
\begin{tabular}{|l|l|l|l|l|l|l|l|l|} \hline
${\rm m}$ & $a_{\rm m}$  & $c_{\rm m}$ & $r_{\rm m}^2$   & $R_{\rm m}^2$ & $K_{\rm m}^2$  &$z^{(1)}_{\rm m}$ & $z^{(2)}_{\rm m}$  & $\delta_{\rm m}$  \\ 
          &               &            & ${\rm GeV^{-2}}$ & ${\rm GeV^{-2}}$ &${\rm GeV^{-2}}$  &                &                     &               \\ \cline{1-9}
0 & 7.8617$\cdot 10^{-3}$& 0.17919             &  0.41493 &0.33040  &  $0.05012$ &     &                & 1.    \\ \cline{1-9} 
1 & 0.14496   & 7.0144$\cdot 10^{-2}$ &0.12531 & 0. &0. & 0.90916    &             &1.5407 \\ \cline{1-9} 
2 & 4.7714 $\cdot 10^{-2}$ & 2.5041$\cdot 10^{-2}$ & 0.10782 &0. &0.  &   0.21016  & 5.7923         & 3.1029   \\ \cline{1-9}
3         &$ -0.22432$     & 1.1516 &$0.027011$   & 0.20426 &0.089174 &  $ 40.533 $& 213.34      &0.65636   \\ \cline{1-9}
{\rm soft}& $ 3.4956\cdot 10^{-3}$ &             &0.10374  & &     &            &             &        \\ \cline{1-9}  
\end{tabular}
\end{center}
\subsection{CD BFKL all flavor eigen-SF}
Here we 
represent the results of numerical solutions for the all flavor eigen-SF
which is the sum of longitudinal and transverse eigen-SF 
$$f_m(Q^2)=f_{\rm Lm}(Q^2)+f_{\rm Tm}(Q^2)$$
 in an analytical form
\beq
f_0(Q^2)=
a_0{R_0^2Q^2\over{1+ R_0^2Q^2 }}
\left[1+c_0\log(1+r_0^2Q^2)\right]^{\gamma_0}\,,
\label{eq:F20}
\eeq

\beq
f_m(Q^2)=a_m f_0(Q^2){1+R_0^2Q^2\over{1+ R_m^2Q^2 }}
\prod ^{m}_{i=1}\left(1-{z\over z^{(i)}_m}\right)\,,\,\, m\geq 1\,,
\label{eq:FN}
\eeq
where $\gamma_0={4\over {3\Delta_0}}$ 
and
\beq
z=\left[1+c_m\log(1+r_m^2Q^2)\right]^{\gamma_m}-1 ,\,\,\,
\gamma_m=\gamma_0 \delta_m\,.
\label{eq:ZFN}
\eeq

 The parameters  tuned to reproduce
 the numerical results for $f_m(Q^2)$ at $Q^2\lsim 10^4\, GeV^2$
are listed in the Table 4.

\vspace{0.5cm}
\begin{center}
Table 4. CD BFKL-Regge structure functions parameters.\vspace{0.45cm}\\

\begin{tabular}{|l|l|l|l|l|l|l|l|l|} \hline
$m$ & $a_m$  & $c_m$ & $r_m^2\,,$ ${\rm GeV^{-2}}$ &
$ R_m^2\,,$ ${\rm GeV^{-2}}$ &
$z^{(1)}_m$ & $z^{(2)}_m$ & $z^{(3)}_m$ &  $\delta_m$ \\ \cline{1-9}
0 & 0.0232  & 0.3261&1.1204&2.6018& & & & 1. \\ \cline{1-9}
1 & 0.2788 &0.1113&0.8755&3.4648&2.4773 &  &  &1.0915 \\ \cline{1-9}
2 & 0.1953 &0.0833&1.5682&3.4824 &1.7706 &12.991&  &1.2450  \\ \cline{1-9}
3 &1.4000  &0.04119 &3.9567 & 2.7706    &0.23585 &0.72853&1.13044   &0.5007 \\ \cline{1-9}
{\rm soft}&0.1077 & 0.0673& 7.0332 & 6.6447 &   &  & &      \\  \cline{1-9}
\end{tabular}
\end{center}
\vspace{0.5cm}

The soft component of the proton structure function is parameterized as follows
\beq
f_ {\rm soft}(Q^2)= 
{a_{\rm soft}R^2_{\rm soft}Q^2\over{1+R^2_{\rm soft} Q^2 }}
\left[1+c_{\rm soft}\log(1+r^2_{\rm soft}Q^2)\right]\,,
\label{eq:FSOFT}
\eeq
with parameters cited in the Table 4.

\vspace{0.5cm}

\newpage

\newpage

{\bf Figure captions}
\begin{enumerate}
\item[{\bf Fig.1}]
The subleading hard-to-rightmost hard and soft-pomeron-to-rightmost hard 
ratio of eigen-structure functions $f_m^{c}(Q^2)/f_0^{c}(Q^2)$  as a function 
$Q^{2}$.

\item[{\bf Fig.2}]
Prediction from CD BFKL-Regge factorization for the  charm structure function 
of the proton $F^{c}_2(x,Q^2)$ as a function of
the Bjorken variable $x_{Bj}$ in comparison with the experimental data from 
 ZEUS  Collaboration \cite{ZEUScc}. The  solid curve is a result of the complete
CD BFKL-Regge expansion, the contribution of the rightmost hard BFKL pole  
(LHA) with $\Delta_{\Pom}=0.4$ is shown by long-dashed line. The dotted curve
in the box for $Q^{2}=4$ GeV$^2$ shows a sum of the rightmost hard BFKL plus
soft-pomeron exchanges (LHSA). The upper long-dashed curve in each box for
$Q^2$ from $1.8$ GeV$^2$ up to $30$ GeV$^2$ corresponds to LHA at $m_c=1.3$ GeV.
At higher $Q^2$ the effect of variation of $m_c$ from $1.5$ to $1.3$ GeV is 
negligible small. 
\item[{\bf Fig.3}]
Predictions from CD BFKL-Regge factorization for open charm photoproduction
cross section $\sigma(\gamma p\to c{\bar c}X)$. The  solid curve is a result 
of the complete
CD BFKL-Regge expansion, the contribution of the rightmost hard BFKL pole  
(LHA) with $\Delta_{\Pom}=0.4$ is shown by long-dashed line. The dotted curve
 shows a sum of the rightmost hard BFKL plus
soft-pomeron exchanges (LHSA). The upper dotted curve corresponds to 
the LHSA with $m_c=1.3$ GeV. The data points are from fixed target 
\cite{DATASIGCC} and H1\&ZEUS HERA \cite{DATAHERACC} experiments. 

\item[{\bf Fig.4}]
The subleading hard-to-rightmost hard and soft-pomeron-to-rightmost hard 
ratio of longitudinal  eigen-structure functions
 $f_{\rm Lm}(Q^2)/f_{\rm L0}(Q^2)$  as a function of
$Q^{2}$.

\item[{\bf Fig.5}]
Prediction from CD BFKL-Regge factorization for the longitudinal structure function 
of the proton $F_L(x_{Bj},Q^2)$ as a function of
the Bjorken variable $x_{Bj}$. The  solid curve is a result of the complete
CD BFKL-Regge expansion, the contribution of the rightmost hard BFKL pole  
(LHA) with $\Delta_{\Pom}=0.4$ is shown by long-dashed line. The dashed curve
 shows the
soft-pomeron contribution.

\item[{\bf Fig.6}]
The subleading hard-to-rightmost hard and soft-pomeron-to-rightmost hard 
ratio of logarithmic derivatives $d_{\rm m}(Q^2)/d_{\rm 0}(Q^2)$
 of the all flavor eigen-SF $d_{m}(Q^2)=\partial f_m/\partial \log Q^2$ 
 as a function of
$Q^{2}$.
\item[{\bf Fig.7}]
Prediction from CD BFKL-Regge factorization for the
 log-derivative 
 of the proton structure function $\partial F_2/\partial \log Q^2$
 as a function of
the Bjorken variable $x_{Bj}$. The  solid curve is a result of the complete
CD BFKL-Regge expansion, the contribution of the rightmost hard BFKL pole  
(LHA) with $\Delta_{\Pom}=0.4$ is shown by long-dashed line. The dashed curve
 shows the
soft-pomeron contribution. Preliminary data by  H1 \cite{LOGH1} and ZEUS \cite{LOGZ}
  are shown by filled and open triangles, respectively.

\end{enumerate}

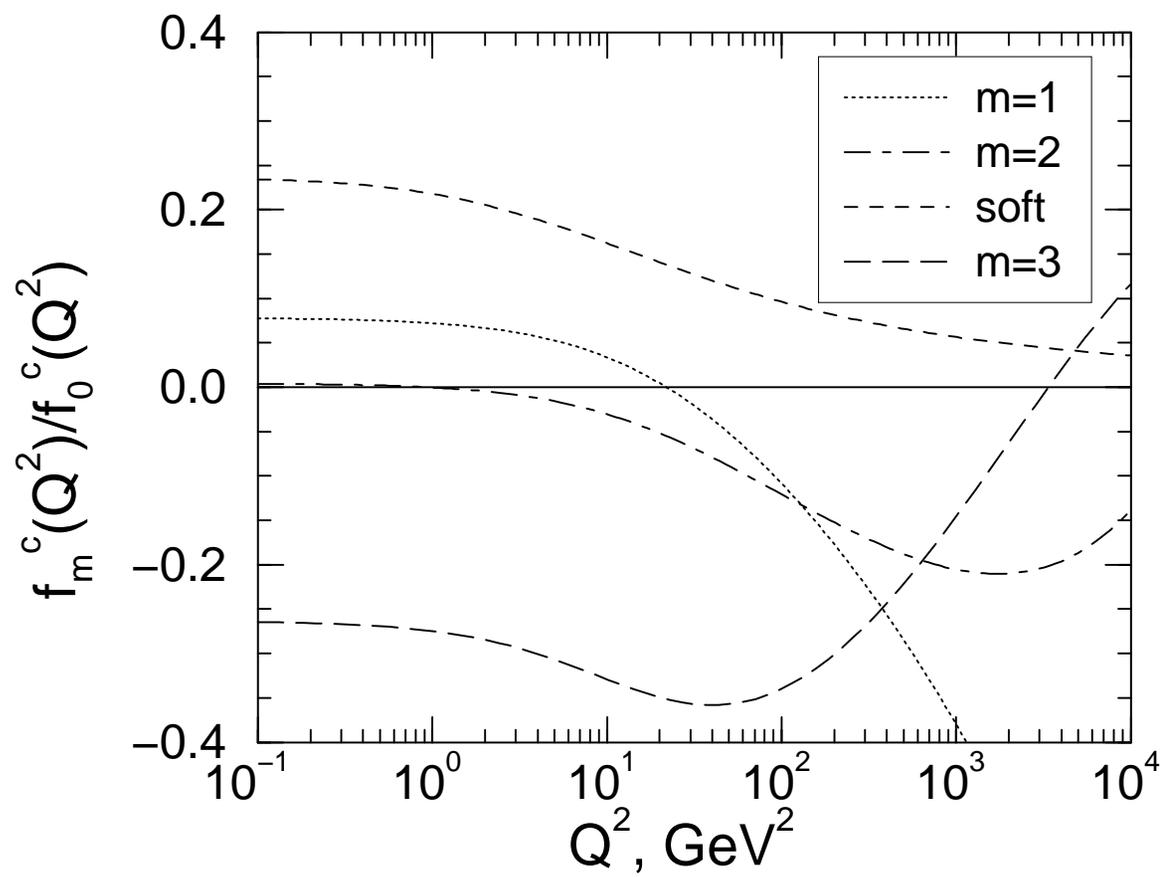
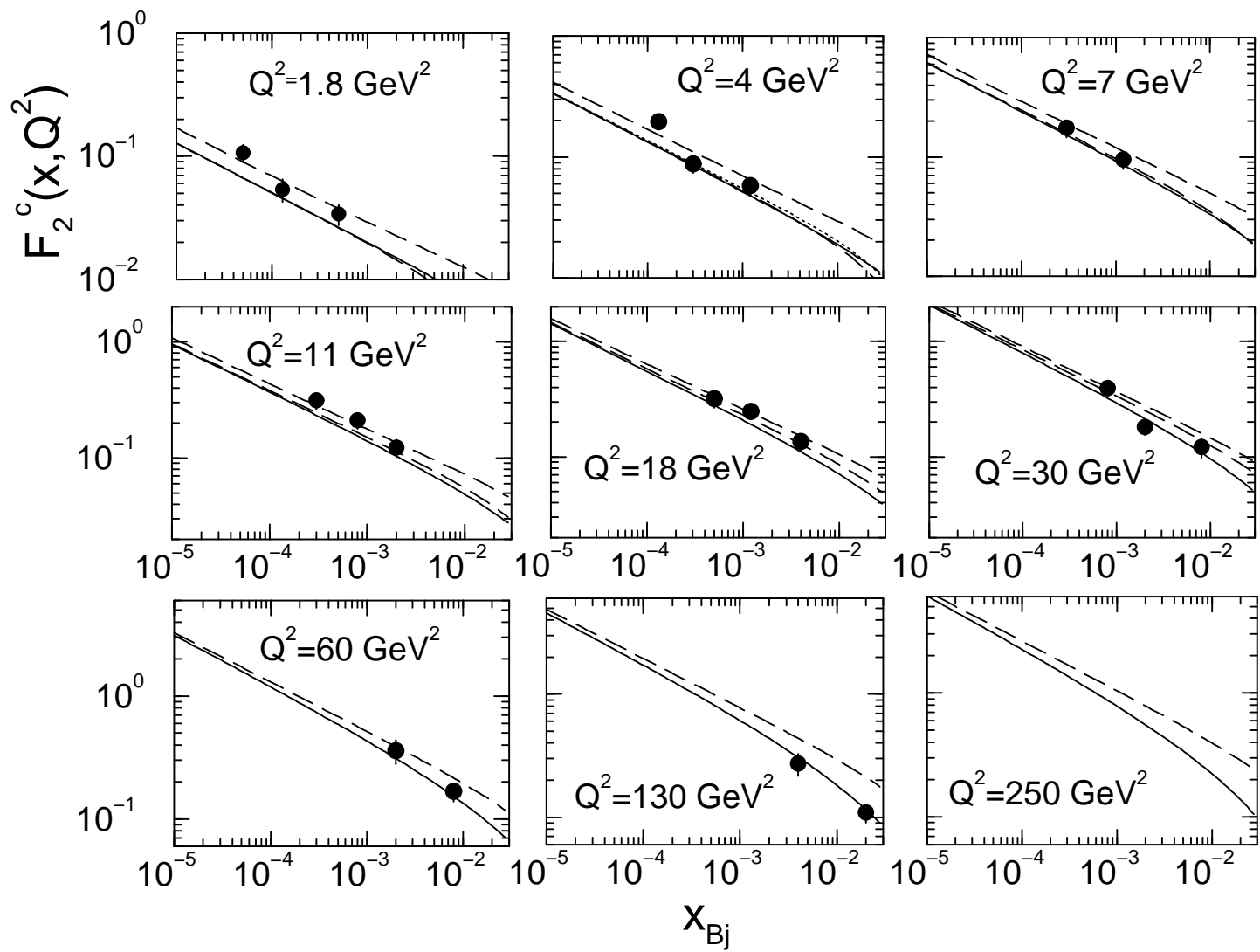
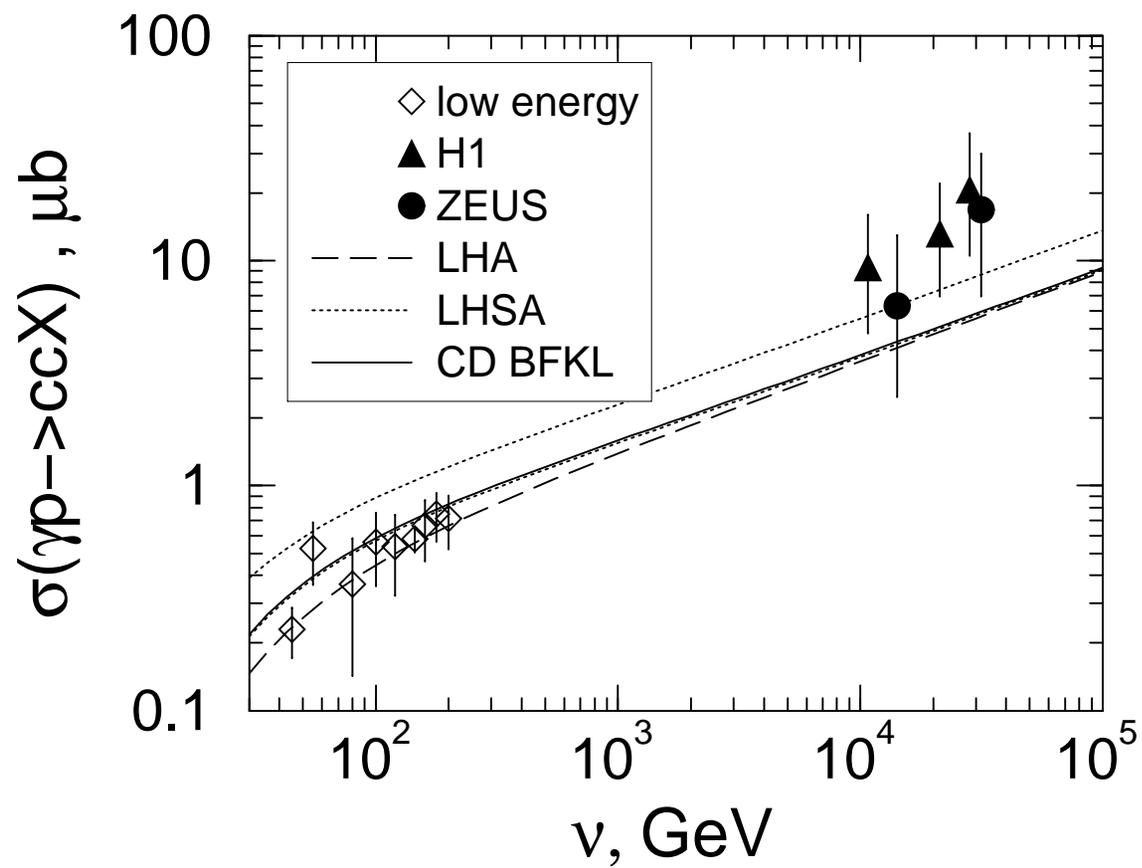
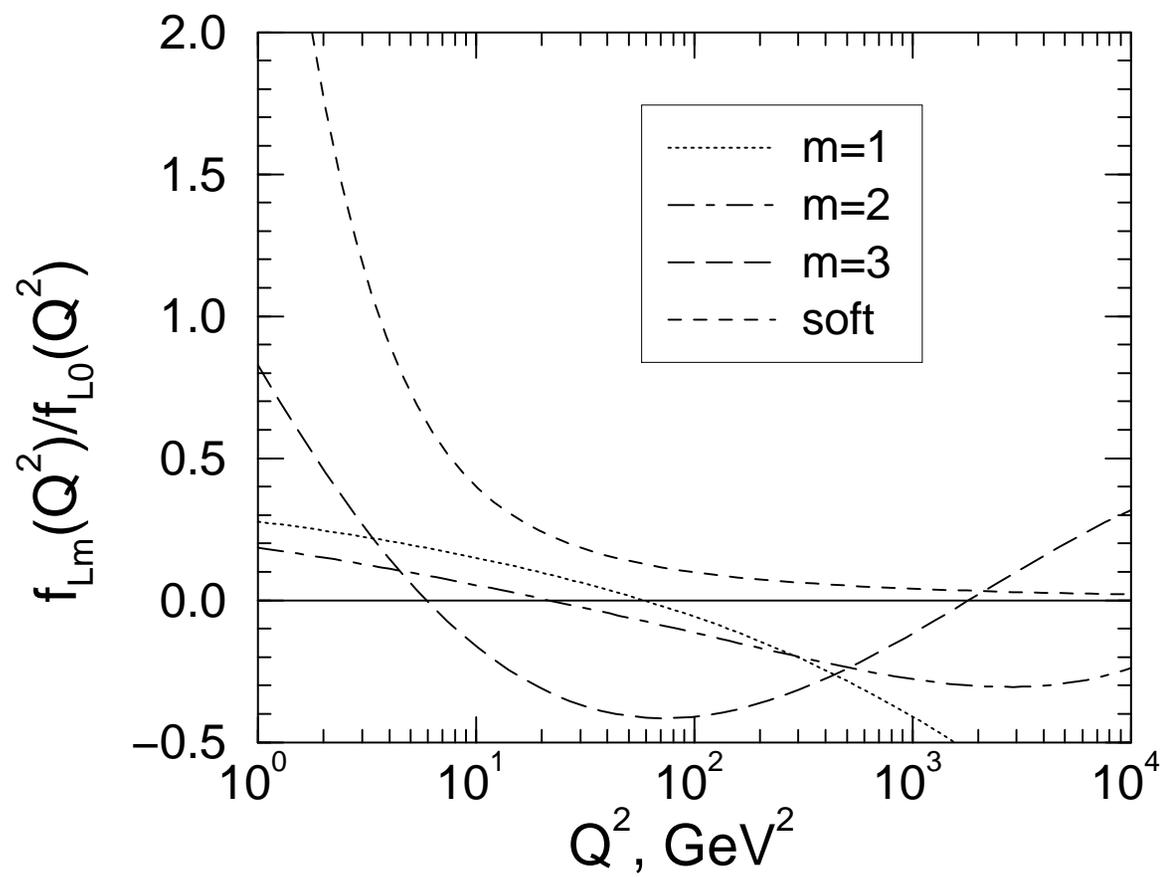
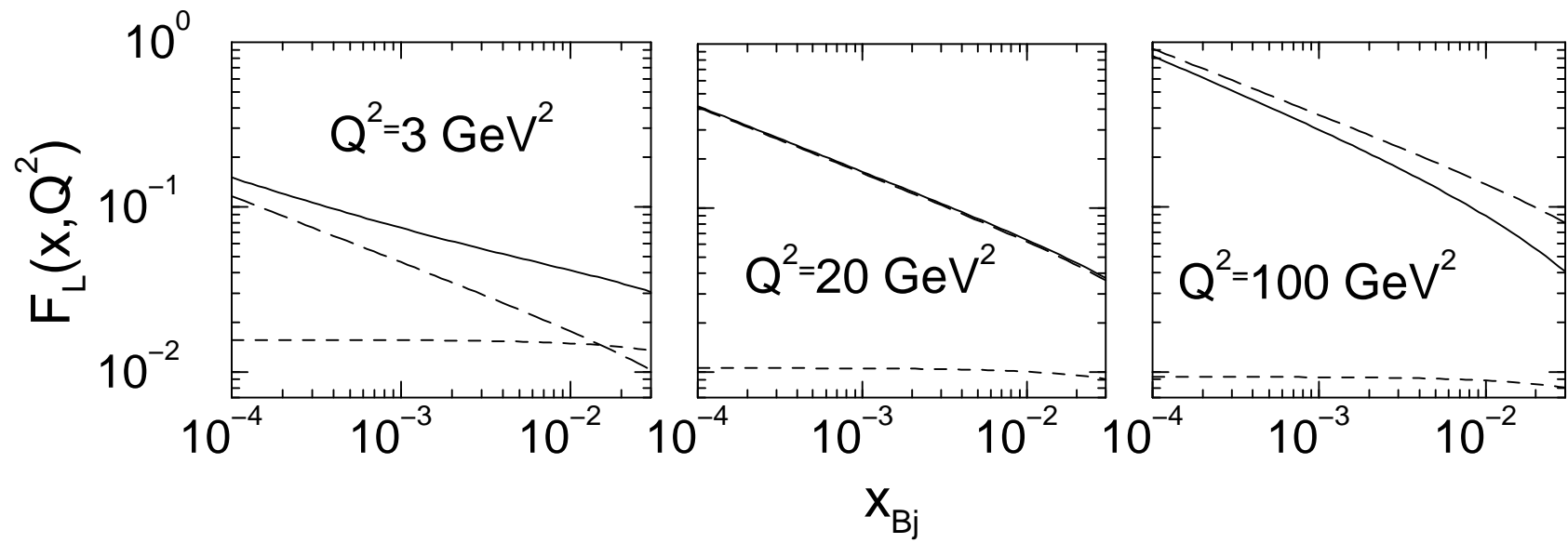
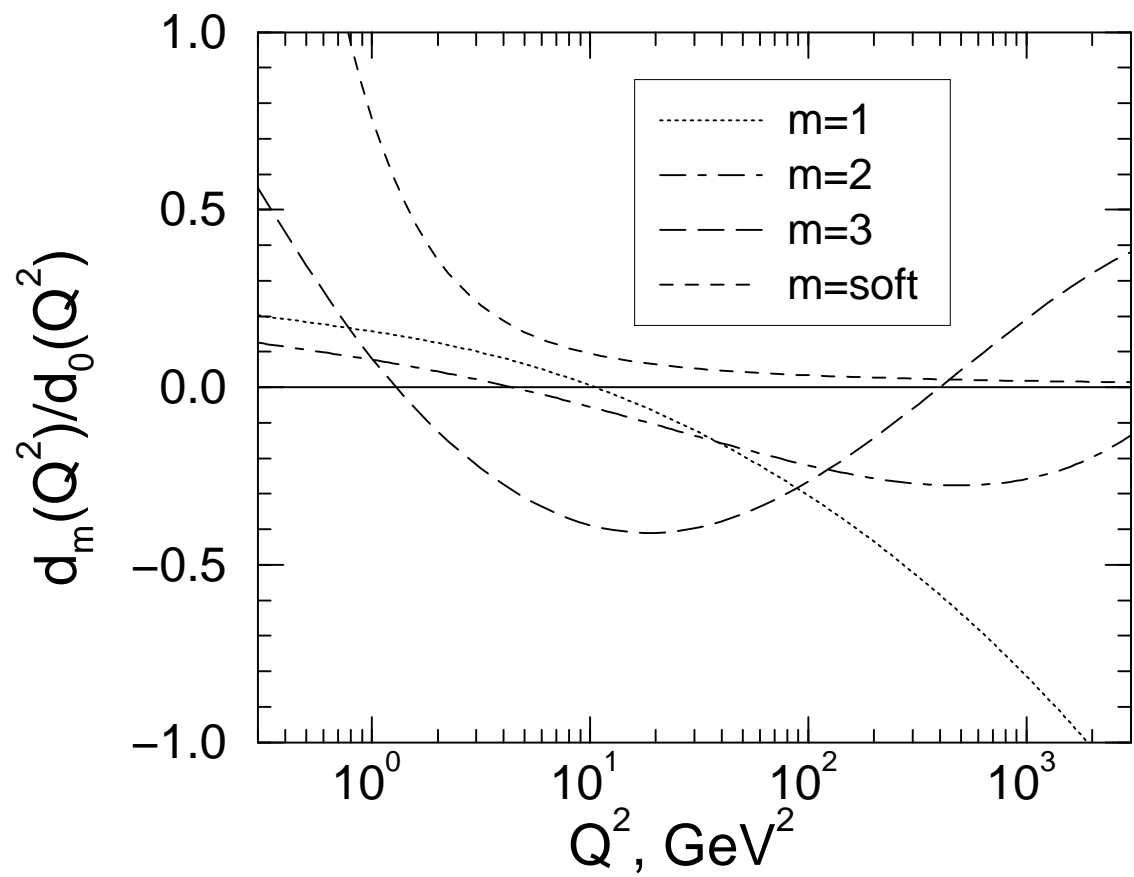
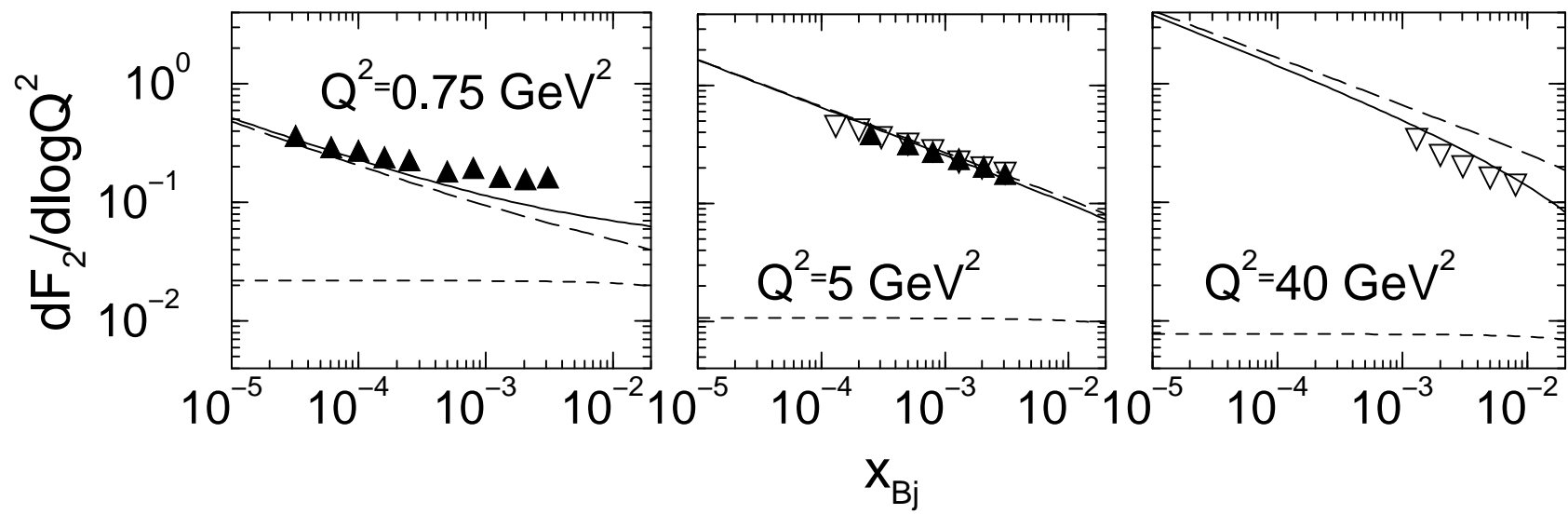
\end{document}